\documentclass[11pt,twoside]{article}
\usepackage{asp2006}
\usepackage{psfig}
\usepackage{epsf}
\usepackage{graphics}
\usepackage{lscape}
\def\edcomment#1{\iffalse\marginpar{\raggedright\sl#1\/}\else\relax\fi}
\reversemarginpar

\begin{document}
\title{Exploring  the  Edge of  the  Stellar  Universe with  Gamma-ray
Observations} \author{F. W.  Stecker} \affil{NASA Goddard Space Flight
Center, Greenbelt, MD, USA} \affil{and Dept. of Physics and Astronomy,
UCLA, Los Angeles, CA, USA} \thispagestyle{plain}

\begin{abstract}
The determination  of the densities of intergalactic  photons from the
FIR to  the UV  produced by stellar  emission and dust  reradiation at
various  redshifts can  provide  an independent  measure  of the  star
formation history of the universe. Using recent {\it Spitzer} and {\it
GALEX} data  in conjunction with other  observational inputs, Stecker,
Malkan and Scully have  calculated the intergalactic photon density as
a function of both energy and redshift  for $0 < z < 6$ and for photon
energies  from 0.003 eV  to the  Lyman limit  cutoff at  13.6 eV  in a
$\Lambda$CDM universe with $\Omega_{\Lambda}  = 0.7$ and $\Omega_{m} =
0.3$.   Their results  are  based on  backwards  evolution models  for
galaxies developed  previously by  Malkan and Stecker.  The calculated
background  SEDs  at  z  =  0  are  in  good  agreement  with  present
observational  data and  limits. The  calculated  intergalactic photon
densities  were  used  to   predict  the  absorption  of  high  energy
$\gamma$-rays in intergalactic space  from sources such as blazars and
quasars,  this  absorption  being  produced  by  interactions  of  the
$\gamma$-rays with intergalactic  FIR-UV photons having the calculated
densities.   The results  are in  excellent agreement  with absorption
features in found in the  very high energy $\gamma$-ray spectra of the
low-$z$ blazars, Mrk 421 and Mrk 501 at $z = 0.03$ and PKS 2155-304 at
$z  = 0.12$.   However,  uncertainties in  the predicted  $\gamma$-ray
absorption  features grow  with redshift.  Actual measurements  of the
spectra  of $\gamma$-ray  sources at  higher redshifts  from detectors
such as the  (soon to be launched) {\it GLAST}  space telescope can be
used to determine intergalactic  photon densities in the distant past,
thereby shedding  light on  the history of  star formation  and galaxy
evolution.
\end{abstract}

\section{Introduction}

Most of this meeting has involved the use of deep astronomical surveys
to study the formation and evolution of galaxies and stars in their 
youngest phases at high redshifts through direct observations of galaxies
at wavelengths
ranging from the radio and submillimeter to the X-ray range. In this
paper, I will discuss a complementary approach which hinges on the
determination of ``historical'' intergalactic IR, optical, and UV photon 
densities produced by the emission of stars and the reradiation of dust
in these young galaxies in the distant past. The energy distribution and 
densities of
these intergalactic photons then become a measure of the {\it total}
star production and population distributions at high redshifts from all
galaxies, visible and obscured, observed and unobserved. 

\begin{figure}

\epsfxsize=8.5cm
\plottwo{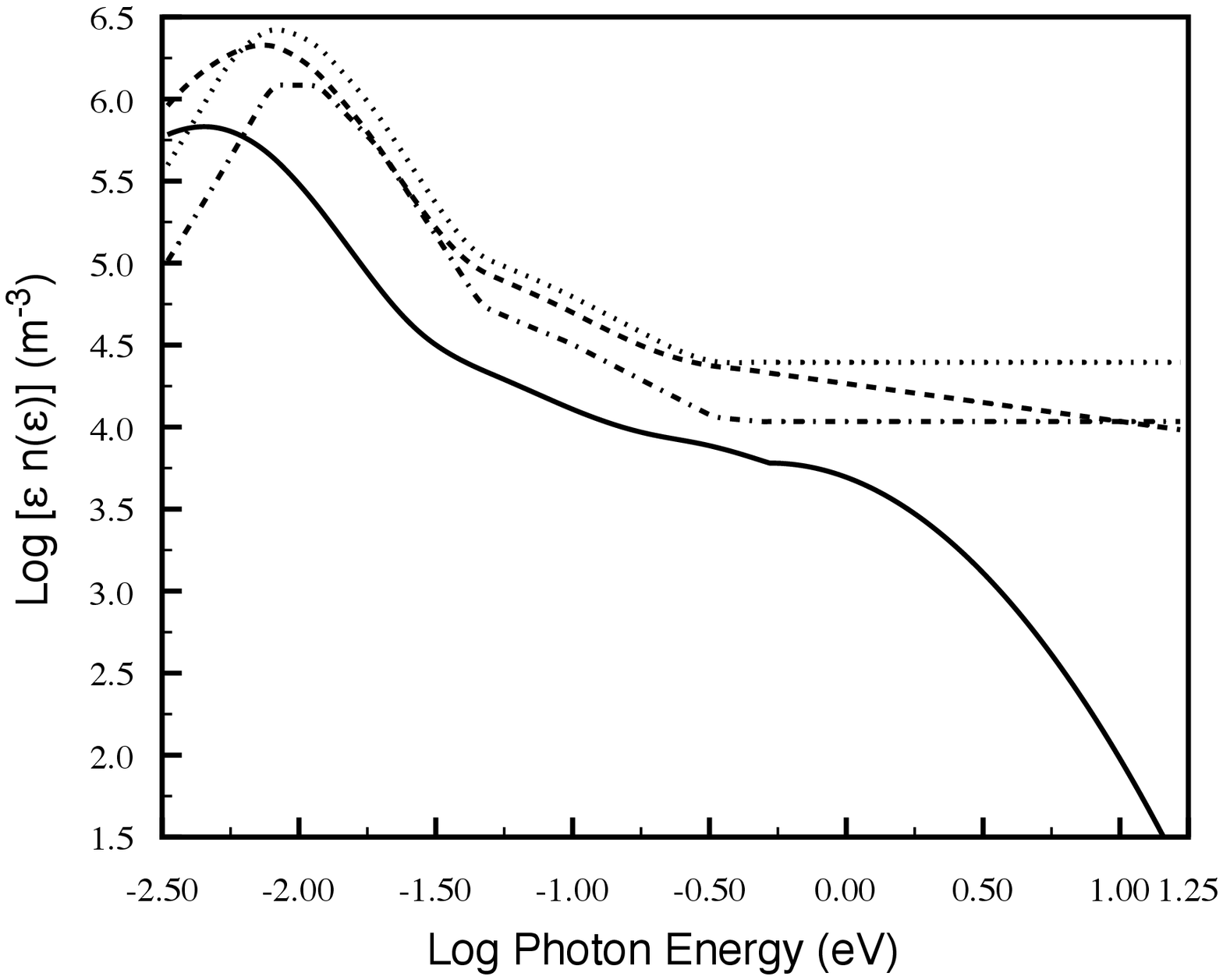}{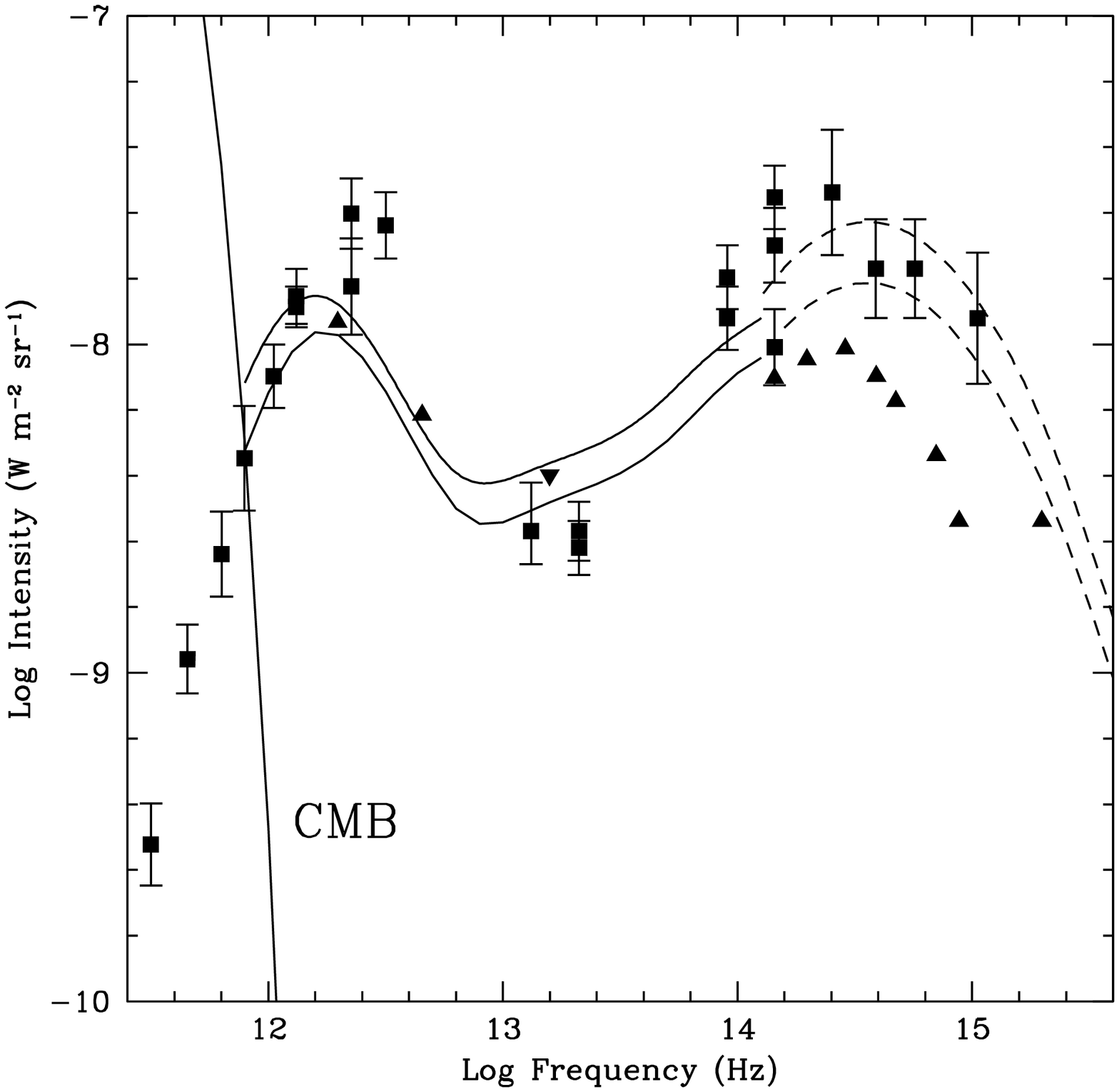}
\caption{The photon density $\epsilon n(\epsilon)$ 
as  a function  of  energy for  various  redshifts based  on the  fast
evolution model. Solid line:  $ z = 0$, dashed line: $ z  = 1$ , dotted
line: $ z = 3$, dot-dashed line: $ z = 5$. (from SMS)}
\label{ste:fig1}

\caption{Spectral  energy   distribution   of  the   diffuse
background  radiation at  $ z  =  0$ from  SMS. Error  bars show  data
points,  triangles  show lower  limits  from  number  counts, and  the
inverted  triangle shows  an upper  limit  from Stecker  and De  Jager
(1997). The  upper and lower solid  lines show the  SMS fast evolution
and baseline  evolution predictions, and  the dotted lines  show their
extensions into the  optical--UV, based on the results  of Salamon and
Stecker (1998).}
\label{ste:fig2}
\end{figure}

\begin{figure}
\epsfxsize=8cm 
\plottwo{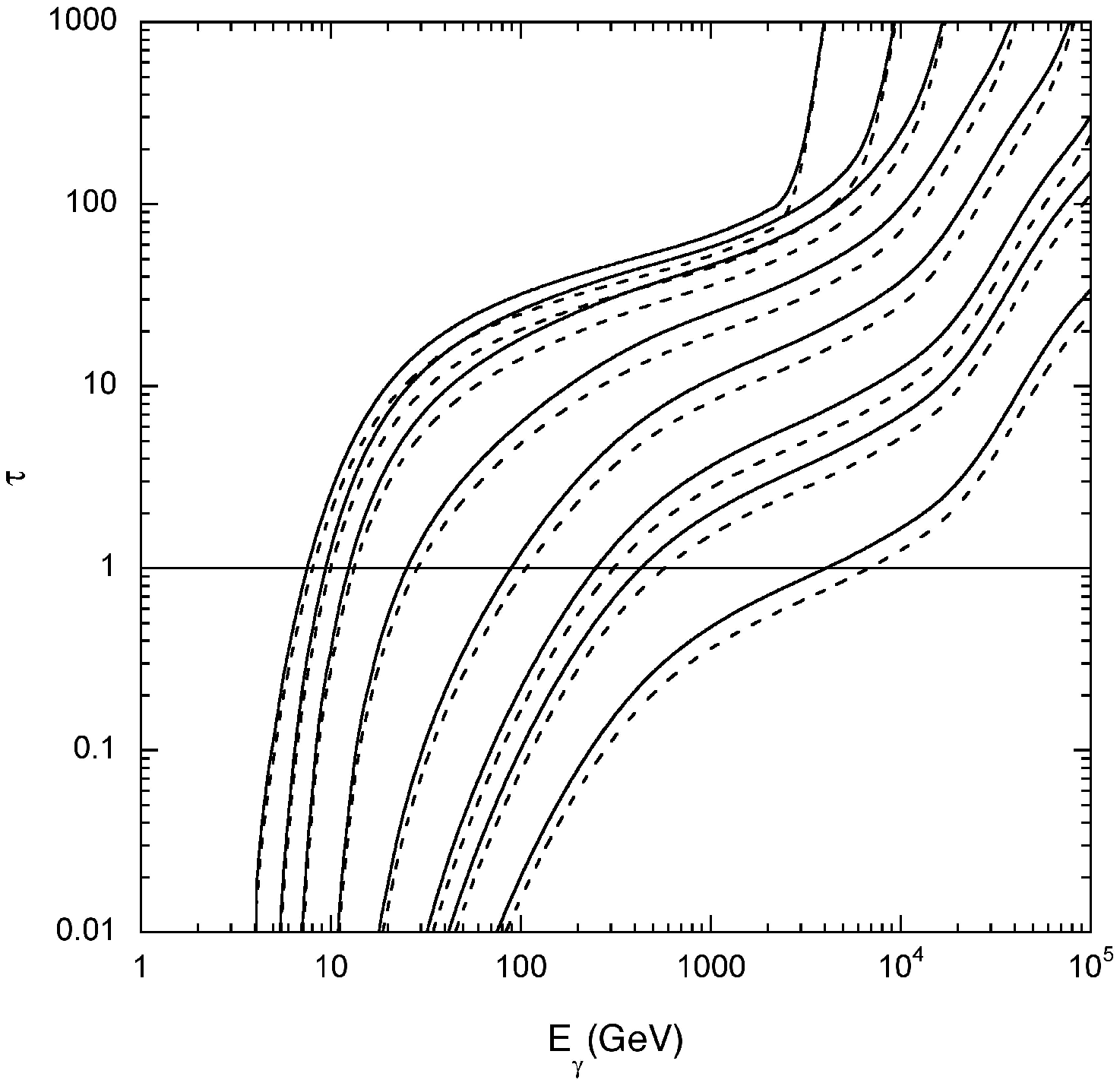}{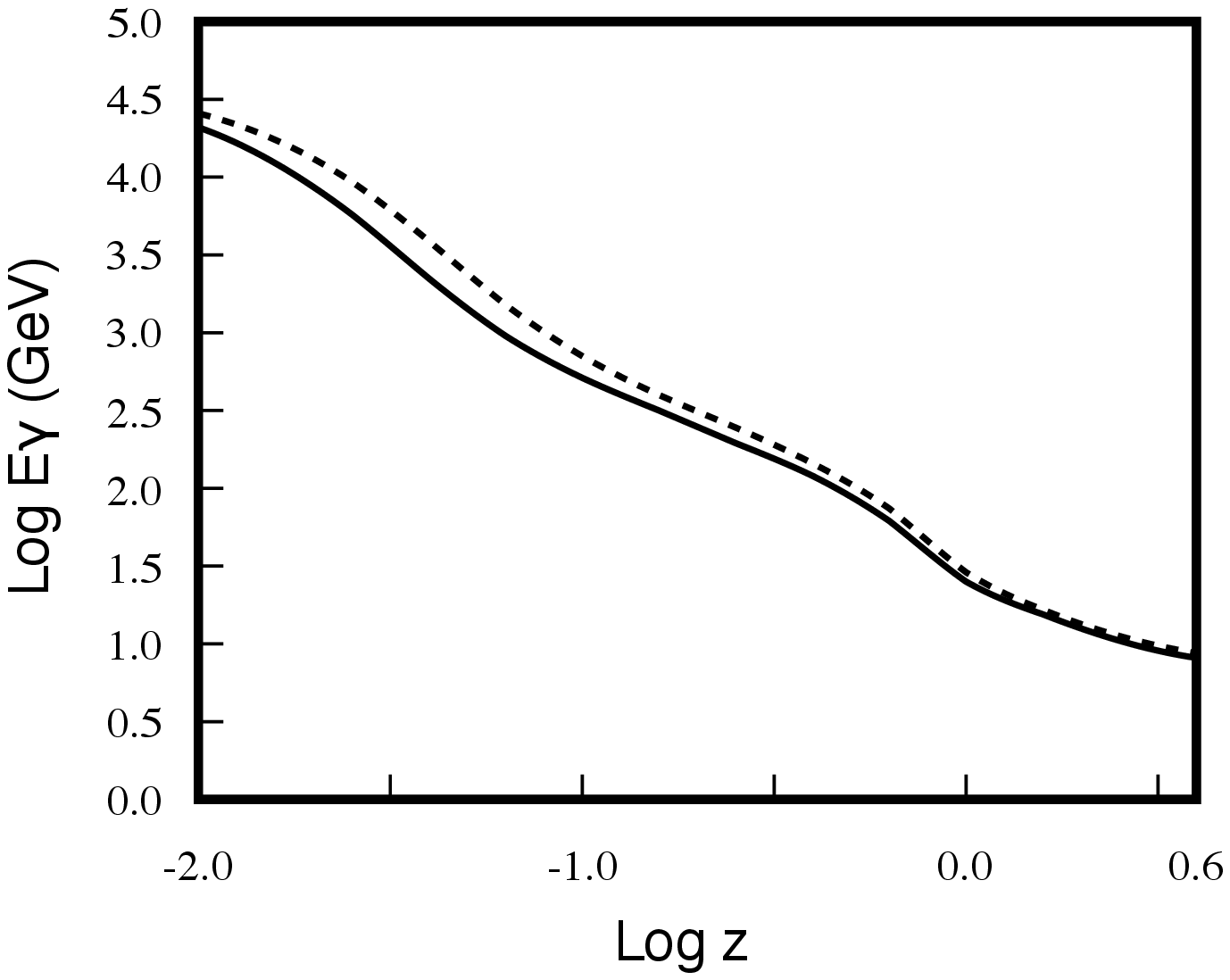}
\caption{The optical  depth of  the universe  to $\gamma$-rays
from interactions  with intergalactic photons, given as a function
of energy for a  family of redshifts from bottom to top of 0.03, 
0.117, 0.2,  0.5, 1.0, 2.0, 3.0 and 5.0. Fast evolution model: 
solid lines, baseline model: dashed lines. (from SMS)}
\label{ste:fig3}

\caption{The critical optical depth $\tau  = 1$ as a function of 
$\gamma$-ray
energy and redshift for the  fast evolution (solid curve) and baseline
(dashed curve) models. Areas to  the right and  above these curves
correspond  to the  region where  the universe  is optically  thick to
$\gamma$-rays. (from SMS)}
\label{ste:fig4}
\end{figure}

\section{Zeroth Order Model Calculations}

As a zeroth order approximation to the expected intergalactic absorption
at high redshifts, we take the results of backward evolution models of
Stecker, Malkan and Scully (2006), hereafter called SMS. These models are 
based on two plausible cases of pure luminosity evolution, {\it viz.}:

\noindent (1)  In the more conservative ``baseline'' scenario,  
all 60$\mu$m galaxy
luminosities  evolved as  $(1+z)^{3.1}$ with their evolution stopped 
at $z_{flat} = 1.4$
and galaxy  luminosities assumed constant (nonevolving)  at the higher
redshifts $1.4 <  z < 6$, with negligible  (assumed zero) emission for
$z  > 6$.   This  later assumption  is  supported by  the recent  {\it
Hubble} deep survey results indicating  that the average  star formation 
rate is  dropping off significantly at a redshift of 6 (Bunker {\it et al.} 
2004; Bouwens {\it et al.} 2005). Independent evidence from luminosity
functions of Lyman break galaxies at redshifts from 3 to 6
indicates a similar decrease in the star formation rate 
(Shimasaku {\it et al.} 2005). However, it is
important to note that the star formation rate for $z  > 6$ is {\it not}
zero and this will modify the results of SMS so that  the predicted
$\gamma$-ray opacity of the high redshift universe will be somewhat
different, depending on the real star formation rate of the universe
at redshifts greater than 6.

\noindent (2) A ``fast  evolution'' scenario where galaxy luminosities
evolved as  $(1+z)^4$ for $0 < z  < 0.8$ and evolved  as $(1+z)^2$ for
$0.8 < z < 1.5$  with no evolution (all luminosities assumed constant)
for for $1.5 < z < 6$  and, again, zero luminosity is assumed for $z >
6$.  This evolution model is  based on the mid-IR luminosity functions
recently determined  out to $  z = 2$  by Perez-Gonzalez {\it  et al.}
(2005).   The ``fast  evolution'' picture  is favored  by  recent {\it
Spitzer}  observations (Le  Floc'h {\it  et al.}  2005, Perez-Gonzalez
{\it et al.}  2005). It provides a better description of the deep {\it
Spitzer}  number   counts  at  70   $\mu$m  and  160$\mu$m   than  the
``baseline'' model.   However, {\it GALEX}  observations indicate that
the evolution of UV  radiation for $0 < z < 1$  may be somewhat slower
and  more  consistent  with   the  ``baseline''  model  within  errors
(Schiminovich {\it et  al.}  2005). Also, the baseline  model fits the
24$\mu$m  {\it  Spitzer} source  counts  more  closely  than the  fast
evolution  model.   The {\it  Spitzer  IRAC}  (Infrared Array  Camera)
counts lie in between the predictions of these two models.

Figure  \ref{ste:fig1}  shows  the  resulting  photon  density  $\epsilon
n(\epsilon)$ as a function of energy at various redshifts for the fast
evolution model. Figure  \ref{ste:fig2} shows the  predicted background SEDs
compared with the  data and empirical limits (see,  {\it e.g.}, Hauser
and Dwek 2001).

\section{The Optical Depth of the Universe to Gamma Rays}

Quantum electrodynamics shows that two photons can annihilate to
produce an electron-positron pair provided that the total energy
in the center of momentum system of the interaction is above the
threshold for making an electron and a positron (Breit and Wheeler
1934). The cross section for this 
interaction peaks close to the threshold energy
(Jauch and Rohrlich 1955). This results in a redshift ``horizon''
for $\gamma$-rays, beyond which the universe is opaque (Stecker 1969;
Fazio and Stecker 1970).
The SMS results on  the optical depth as a function  of energy for various
redshifts   out   to  a   redshift   of   5   are  shown   in   Figure
\ref{ste:fig3}. Figure  \ref{ste:fig4} shows the  energy-redshift relation
giving an  optical depth $\tau =  1$ based on the  SMS calculations of
$\tau (E_{\gamma}, z)$. 

As is shown in Figure \ref{ste:fig3}, for $\gamma$-ray sources
at the higher redshifts there  is a steeper energy dependence of $\tau
(E_{\gamma})$ near the energy where $\tau  = 1$. 
This effect is caused  by the sharp drop in the UV photon
density  at  the  Lyman   limit.  There will thus be a
sharper absorption cutoff in the multi-GeV $\gamma$-ray spectrum of 
sources at high redshifts than in the TeV spectra of more nearby
$\gamma$-ray sources. It is important
to note that {\it the exact position of this cutoff in energy is directly
related to the $z$-dependence of the star formation rate at high redshifts.}

\subsection{Implications for {\it GLAST} }

Because absorption cutoffs in the spectra of blazars
at the higher redshifts lies in the multi-GeV range, {\it GLAST}, the Gamma
Ray Large Space  Telescope ({\tt http://glast.gsfc.nasa.gov}), to be launched
in the fall of 2007, will be able to make measurements of such features
and thus probe the early star formation  rate (Chen, Reyes
\&  Ritz 2004).   {\it  GLAST} will  be able  to detect
blazars at $z \sim 2$ at multi-GeV energies, the critical energy
range for the expected sharp absorption cutoffs in high redshift
$\gamma$-ray sources as shown in Figure \ref{ste:fig3}.
Thus, $\gamma$-ray observations by {\it GLAST} can complement the  
deep galaxy surveys and help determine the
redshift when significant star formation began.
In fact,  {\it GLAST} need  not have to  detect $\gamma$-ray sources  at high
redshifts in  order to aquire  information about the evolution  of 
intergalactic photon fluxes. If the diffuse  $\gamma$-ray background  
radiation is  from unresolved
blazars  (Stecker  \&  Salamon   1996),  a  hypothesis  which  can  be
independently  tested by {\it  GLAST} (Stecker  \& Salamon  1999), the
effects of $\gamma$-ray absorption will  steepen the spectrum of this radiation
at $\gamma$-ray energies above $\sim 10$  GeV (Salamon \& Stecker 1998).

\end{document}